\documentclass[superscriptaddress,prl,twocolumn,showpacs]{revtex4}
\usepackage{times,graphicx,amsmath,amssymb}

\def\kbar{\protect\@kbar}
\def\@kbar{%
\relax \bgroup
\def\@tempa{\hbox{\raise.73\ht0
\hbox to0pt{\kern.25\wd0\vrule width.5\wd0
height.1pt depth.1pt\hss}\box0}}%
\mathchoice{\setbox0\hbox{$\displaystyle k$}\@tempa}%
{\setbox0\hbox{$\textstyle k$}\@tempa}%
{\setbox0\hbox{$\scriptstyle k$}\@tempa}%
{\setbox0\hbox{$\scriptscriptstyle k$}\@tempa}%
\egroup}

\newcommand{\ps}{\ps}

\input epsf

\begin{document}

\title{Decoherence as a Probe of Coherent Quantum Dynamics}

\author{Michael B. d'Arcy}
\affiliation{Clarendon Laboratory, Department of Physics,
University of Oxford, Parks Road, Oxford, OX1 3PU, United Kingdom}

\author{Rachel M. Godun}
\affiliation{Clarendon Laboratory, Department of Physics,
University of Oxford, Parks Road, Oxford, OX1 3PU, United Kingdom}

\author{Gil S. Summy}
\affiliation{Clarendon Laboratory, Department of Physics,
University of Oxford, Parks Road, Oxford, OX1 3PU, United
Kingdom}\affiliation{Department of Physics, Oklahoma State
University, Stillwater, Oklahoma, 74078-3072, USA}

\author{Italo Guarneri}
\affiliation{Universit\`{a} degli Studi dell'Insubria, Via
Valleggio 11, I-22100 Como, Italy} \affiliation{Instituto
Nazionale per la Fisica della Materia, Unit\`{a} di Milano, Via
Celoria, I-20133 Milano, Italy} \affiliation{Instituto Nazionale
di Fisica Nucleare, Sezione di Pavia, Via Bassi 6, I-27100 Pavia,
Italy}

\author{Sandro Wimberger}
\affiliation{Instituto Nazionale per la Fisica della Materia,
Unit\`{a} di Milano, Via Celoria, I-20133 Milano, Italy}
\affiliation{Instituto Nazionale di Fisica Nucleare, Sezione di
Pavia, Via Bassi 6, I-27100 Pavia, Italy}
\affiliation{Max-Planck-Institut f\"{u}r Physik komplexer Systeme,
N\"{o}thnitzer Str.\ 38, D-01187 Dresden, Germany}

\author{Shmuel Fishman}
\affiliation{Physics Department, Technion, Haifa IL-32000, Israel}

\author{Andreas Buchleitner}
\affiliation{Max-Planck-Institut f\"{u}r Physik komplexer Systeme,
N\"{o}thnitzer Str.\ 38, D-01187 Dresden, Germany}

\date{\today}

\begin{abstract}
The effect of decoherence, induced by spontaneous emission, on the
dynamics of cold atoms periodically kicked by an optical lattice
is experimentally and theoretically studied. Ideally, the mean
energy growth is essentially unaffected by weak decoherence, but
the resonant momentum distributions are fundamentally altered. It
is shown that experiments are inevitably sensitive to certain
nontrivial features of these distributions, in a way that explains
the puzzle of the observed enhancement of resonances by
decoherence [Phys.\ Rev.\ Lett.\ {\bf 87}, 074102 (2001)]. This
clarifies both the nature of the coherent evolution, and the way
in which decoherence disrupts it.
\end{abstract}


\maketitle

The theory of coherent quantum transport in periodic potentials is
basic to solid state physics, and to our understanding of various
conductance phenomena in crystal lattices. After succeeding in
isolating and manipulating single quantum objects such as ions or
atoms, quantum opticians, as well as meso- and nanoscientists,
have now started to build extended structures of atoms or ions of
increasing complexity. A natural way of doing so is to arrange
one, two, or three dimensional regular arrays of (cold or
ultracold) atoms in optical lattices \cite{greiner2002}, which
then can be considered as faithful realizations of strongly
idealized, fundamental models of solid state theory. Beyond
illustrating such theoretical models under clean and virtually
perfectly controlled laboratory conditions, these experiments
often also hold unexpected surprises (due to apparently innocent,
real-life modifications of the original model), and promise highly
rewarding technical applications in the future. Proposals that
suggest using optical lattices for quantum information processing
\cite{jaksch99} are but one example of this.

In all such respects, the impact of noise and decoherence is a
crucial issue \cite{dittrich86,ammann98,buchleitner2002}, because
decoherence is expected to impair manifestly quantum phenomena.
The present Report addresses a striking, seeming violation of this
rule, which was experimentally observed with kicked cold atoms
subjected to a pulsed, one-dimensional, spatially periodic optical
lattice \cite{darcy2001b}. Here, a peculiar type of coherent
quantum transport, called `quantum resonance'
\cite{izrailev79,WGF2003} is theoretically predicted for kicking
periods rationally related to the propagation time of kicked atoms
across the lattice constant. The mean kinetic energy of an atomic
ensemble is then predicted to increase linearly with time, in
sharp contrast to the behavior predicted for non-resonant values
of the kicking period, where it saturates (in the process of
`dynamical localization' \cite{fishman93}, closely analogous to
Anderson localization in one-dimensional disordered solids
\cite{anderson58}). In previous experiments, enhanced transport
was indeed observed for the lowest-order resonances
\cite{darcy2001b,oskay2000}. When decoherence was added to the
experiment by controlled spontaneous emission (SE), the energy
growth at resonance was found to be significantly faster than
could be accounted for by the heating effect of momentum transfer
due to SE \cite{darcy2001b,darcy2001a}. This looks like incoherent
magnification of a purely coherent, and non-classical, phenomenon.

In this Report we show how this counter-intuitive effect of
decoherence can be resolved by inspection of the full atomic
momentum distributions instead of merely their mean square (i.e.,
kinetic energy) values, as were considered in
Ref.~\cite{darcy2001b}. Decoherence then acts as expected: it
destroys the coherent dynamics underlying the quantum resonances
and, in particular, certain nontrivial features of the momentum
distributions \cite{WGF2003}. It is precisely this latter fact
that produces the surprising enhancement of the mean energy values
observed in the experiment \cite{darcy2001b,WGF2003}.

Our experimental system \cite{darcy2001a} is a realization of the
paradigmatic Kicked Rotor (KR) model \cite{fishman93,graham92},
extensively used in investigations of classical chaotic dynamics
and its quantum counterpart \cite{stockmann99}. After trapping and
cooling in a magneto-optic trap (MOT), about $10^{7}$ cesium atoms
are released and, falling freely under gravity, are exposed to
pulses from a vertical standing wave of off-resonant laser light.
This is red-detuned from the $6^{2}\mbox{S}_{1/2} \rightarrow
6^{2}\mbox{P}_{1/2}$, $(F=4 \rightarrow F'=3)$ D1 transition by
$\delta_{L}=2\pi \times 30$\thinspace GHz, and has a wavelength
$\lambda_{L} = 894.7$\thinspace nm. On release, the atomic
temperature is $5\mu$K, corresponding to a Gaussian momentum
distribution with full width at half maximum (FWHM) 12$\hbar
k_{L}$, where $k_{L}=2\pi/\lambda_{L}$. The duration of each
(square) pulse is $t_{p}=500$\thinspace ns, and the peak intensity
in the standing wave is $\simeq 5 \times 10^{4}$ mW/cm$^{2}$. Due
to the ac Stark shift, these pulses result in
$\delta$-function-like applications of a sinusoidal potential,
with spatial period $\lambda_{L}/2$. Classically, the maximum
impulse that this can impart is $\hbar G\phi_{d}$, where $\phi_{d}
= \Omega^{2}t_{p}/8\delta_{L}$, and $\Omega$ is the Rabi frequency
of the atoms at the intensity maxima of the light field. Quantum
mechanically, it imparts momentum to the atoms in integer
multiples of $\hbar G$, where $G=2k_{L}$.  Both the density
distribution of the trapped atoms and the standing light wave
intensity profile are Gaussian, each with FWHM $1$\thinspace mm,
so the mean value of $\phi_{d}$ as experienced by the atomic
ensemble is $\simeq 0.8\pi$. The standing wave passes through a
voltage-controlled crystal phase modulator which can shift the
position of the standing wave between consecutive pulses so that
it effectively cancels the effect of gravity. Thus in the rest
frame of the atomic ensemble the standing wave appears to be
stationary, yielding KR dynamics with kicking period $T$ ($\gg
t_{p}$), despite the presence of gravity. Decoherence can be
introduced by inducing SE in the atoms through application of an
additional $2\mu$s pulse of laser light after each of the kicking
pulses. This light is $60$\thinspace MHz red-detuned from the
$6^{2}\mbox{S}_{1/2}\rightarrow 6^{2}\mbox{P}_{3/2}$, $(F=4
\rightarrow F''=5)$ D2 transition. The intensity of the pulse can
be controlled so that the mean number of SE per atom per pulsing
cycle, ${\overline n}_{\mbox{\tiny SE}}$, can be varied
continuously. Finally, after application of the pulses, the atoms
fall through a sheet of light resonant with the D2 transition,
located $0.5$m below the point of release. This allows us to
determine their momentum distribution by a time-of-flight (TOF)
technique, with a resolution of $\simeq \hbar G$.

In the absence of SE, the Hamiltonian that generates the time
evolution of the atomic wave function may be written in the
following dimensionless form
\begin{equation}
\hat{H}(t)=\frac{\hat{p}^2}{2}+\phi_{d}\cos(\hat{x})\sum_{m=-\infty}^{+\infty}\delta
(t-m\tau) \, , \label{eq:ham}
\end{equation}
where $p$ is momentum in units of $\hbar G$, $x$ is position in
units of $G^{-1}$, and $M$ is the mass of the atoms. The units of
energy and time are then $\hbar^2G^2/M$ and $M/\hbar G^2$,
respectively. In such units the kicking period is $\tau=\hbar
G^2T/M$. This Hamiltonian is very close to that of the well-known
$\delta$-kicked rotor, with the sole (but important, as discussed
below) difference that our cold atoms are moving along a line
rather than in a circle. Notwithstanding this, the dynamics of the
atoms reflect characteristic properties of the quantum KR (whose
corresponding classical phase space is mixed, consisting of
regular and chaotic components, for non-vanishing kicking strength
\cite{lichtenberg92}). The nature of the quantum transport
sensitively depends on the parameter $\tau$. If $\tau=4\pi r/q$,
with $r,q$ integers, then the kicking period is rationally related
to the propagation time of kicked atoms across the lattice
constant, and quantum transport is typically enhanced by quantum
resonance \cite{izrailev79}. If $\tau/4\pi$ is sufficiently
irrational, then transport is inhibited by quantum interference,
i.e., by dynamical localization \cite{fishman93}.

\begin{figure}
\centering
\includegraphics[width=8.5cm]{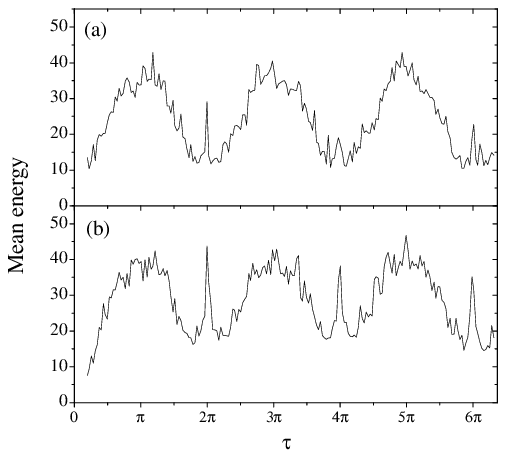}
\caption{Experimental values of the mean energy of the atomic
ensemble after 30 kicks, as $\tau$ is varied from $0.19\pi$ to
$6.31\pi$ (i.e., $6.5\ {\rm \mu s}\leq T \leq 210.5 {\rm \mu s}$)
in (a) the absence, and (b) the presence of induced spontaneous
emission, with ${\overline n}_{\mbox{\tiny SE}}\simeq 0.14$.}
\label{Fig:tscan1}
\end{figure}

A conceptually simple way of experimentally testing this
theoretical picture is to measure as a function of $\tau$ the mean
kinetic energy of the atoms, henceforth to be referred to as `mean
energy', after a fixed interaction time of $N$ kicks. The result
of such a measurement at $N=30$ in the absence of SE is shown in
Fig.~\ref{Fig:tscan1}(a), with $0.19\pi \leq \tau \leq 6.31\pi$.
The mean energies were extracted from a finite momentum window
($-60\leq p \leq 60$), and low-amplitude noise in the
time-of-flight signal was eliminated by imposing a signal
threshold estimated from the background noise level at high
momenta. Figure~\ref{Fig:tscan1}(a) shows an underlying smooth,
periodic dependence on $\tau$. This reflects the $\tau$-dependence
of the localization length of the dynamically localized atomic
sample \cite{shepelyansky87}. The structure superimposed on the
basic periodic variation has particularly narrow peaks at
$\tau=2\pi,4\pi,6\pi$. These are the main quantum resonances, with
$q=1,r=1$ and $q=2,r=1,3$. Higher-order ($q \geq 3$) resonances
were not unambiguously resolved within the given observation time.
The present Report is therefore focused on the main resonances,
and specifically at the surprising way in which they react to
decoherence, as shown in Fig.~\ref{Fig:tscan1}(b). The mean energy
growth at the resonances is clearly enhanced. In other words,
resonant transport, which is due to constructive quantum
interference, appears to be stabilized by decoherence (while the
dynamical localization away from resonance is barely affected,
confirming that we are in the regime of weak decoherence
\cite{dittrich86}). The apparent inconsistency of these
experimental observations with what seems theoretically reasonable
is resolved as follows.

\begin{figure}
\centering
\includegraphics[width=8.5cm]{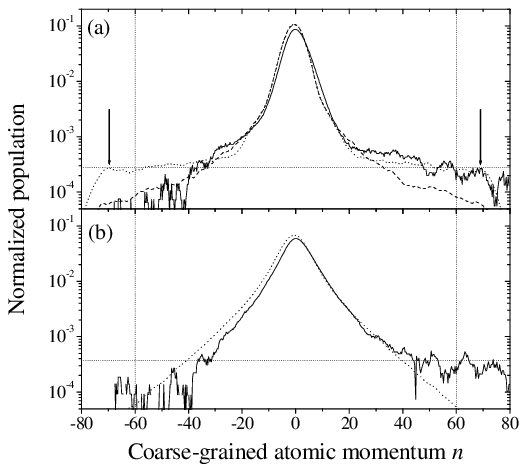}
\caption{Normalized experimental (solid line) and coarse-grained
numerical (dotted) momentum distributions after $N=30$ kicks at
the quantum resonance $\tau \simeq 2\pi$ ($T=66.5\mu$s) in (a) the
absence, and (b) the presence of induced spontaneous emission.
Experimentally, ${\overline n}_{\mbox{\tiny SE}}= (0.14\pm0.04)$,
while ${\overline n}_{\mbox{\tiny SE}}=0.1$ numerically. The
dashed curve in (a) is the asymptotic distribution $P_{s}(n)$, as
given in Eq.~(\ref{eq:fourdis}). The arrow labels in (a) indicate
the ballistic wings, whose momentum varies linearly with $N$. Note
the slight asymmetry in the experimental distributions around
$n=0$, due to non-ideal aspects of the realization. The fainter
dotted lines show the signal threshold and momentum cuts imposed
on the experimental data when calculating mean energies.}
\label{Fig:quantres}
\end{figure}

An atom periodically kicked in space and time is described by a
wave packet $\psi(x)$ composed of $2\pi$-periodic Bloch states
$\psi_{\beta}(x)$, that is,
\begin{equation}
\psi(x)=\int_0^1d\beta\exp(i\beta x)\psi_{\beta}(x)\, ,
\label{eq:bloch}
\end{equation}
where $\beta$ is the quasimomentum. In our units, it is given by
the fractional part of the momentum $p = n + \beta$ ($n \in
\mathbb{Z}$). It is conserved in time, so the different Bloch
states in (\ref{eq:bloch}) evolve independently of one another,
and their momenta only change by integers. Under the resonance
condition $\tau=4\pi r/q$, a special situation occurs for a
specific, discrete subclass of values of $\beta$. Besides being
periodic in coordinate space, the one-period evolution (Floquet)
operator is then also periodic in momentum space, with the integer
period $q$. This happens when $\beta = m/2r$, $0\leq m<2r$, $m$
integer. The amplitudes of such waves at momentum states separated
by $q\hbar G$ (in physical units) exactly rephase after each kick
\cite{darcy2001b}; here we specialize to $q=1,2$. This rephasing
is analogous to the Talbot effect in optics, so we speak of these
resonances as occurring at rational multiples of the half-Talbot
time $T_{1/2}= 2\pi M/\hbar G^{2} = 66.7\mu$s (for which $\tau =
2\pi$). In much the same way as spatial periodicity enforces
ballistic motion in physical space, the momentum periodicity which
holds for special values of $\beta$ (i.e., $\beta=1/2$ for $q=2$,
and $\beta=0,1/2$ for $q=1$) enforces ballistic propagation of the
corresponding states {\it in momentum space}; thus their energy
grows quadratically in time. The remaining Bloch components of the
original wavepacket, with $\beta$ not in the `resonant' class,
undergo a quasiperiodic energy exchange with the driving field,
leading to a finite spread of the associated ($\beta$-dependent)
momentum distribution for all times. Upon incoherently averaging
over the continuous set of quasimomenta which constitute the
atomic ensemble, there is competition between quasi-periodic and
ballistic propagation, and as $N$ increases the values of $\beta$
that populate the ballistic growth must match more closely the
ideal resonant values. On the one hand, this leads to {\em
linear}\/ growth of the total mean energy, $E\approx
\phi_{d}^{2}N/4$. On the other hand, a {\em stationary}\/ momentum
distribution $P_{s}(n)=\lim_{N\to\infty}P(n,N)$ \cite{WGF2003},
given by
\begin{equation}
\label{eq:fourdis} P_{s}(n) = \sum\limits_{n'} h(n')
\int_{-\pi}^{\pi}\frac{d\xi}{2\pi}\int_0^{2\pi}
\frac{d\alpha}{2\pi}\;J^2_{n-n'}(f(\xi,\alpha))\; ,
\end{equation}
emerges, where $P(n,N)$ is the coarse-grained momentum
distribution of the ensemble after $N$ kicks. The coarse-graining
is on the scale of unity ($\hbar G$ in physical units) so as to
yield a distribution in $n$, which is consistent with the
finite-size binning of the experimentally detected momentum
distribution. In Eq.~(\ref{eq:fourdis}), $h(n')$ is the initial
(assumed smooth) momentum distribution, $f(\xi,\alpha) =
\phi_{d}\sin(\xi)\csc(\alpha)$, $\xi = \pi(2\beta - 1)$, and
$J_{n-n'}$ is a Bessel function of first kind and order $n-n'$.
The asymptotic distribution $P_{s}(n)$, shown in
Fig.~\ref{Fig:quantres}(a), is attained because the phases
$\alpha$ of the nonresonant Bloch components of the original
wavepacket, accumulated under the action of the time evolution
operator, are effectively averaged.

For finite times, $P(n,N)$ exhibits a narrow, stationary peak
centered around $n=0$, algebraic decay $\propto n^{-2}$ over
intermediate momenta, and `ballistic wings' due to the
almost-resonant $\beta$ values, which move to higher momenta
linearly in time. It is important to note that the central peak is
{\em narrower}\/ than the exponential distribution observed in
dynamically localized atomic ensembles, and that the linear energy
growth is observed at all times, in spite of the onset of the
stationary distribution. There is no inconsistency here: the
asymptotic limit to which the distribution tends has an $n^{-2}$
fall-off at large $n$, and hence has an infinite mean energy
(i.e., a divergent 2nd moment). The ballistic wings, also
experimentally observed in Ref.~\cite{oskay2000}, dominate the
theoretically computed mean energy growth. As the wings are fed by
the resonant-$\beta$ subclass, the resonant energy growth is
ultimately due to conservation of the quasimomentum $\beta$ in the
kicking process.

Experimental detection of these wings in the final atomic momentum
distribution is extremely difficult, for several reasons. The most
important of these for our present discussion is that the wings
must not have moved beyond the cutoffs imposed by the
signal-to-noise ratio. Though relatively small in terms of
population, this experimental loss from the ballistic wings leads
to a mean energy which is significantly less than the theoretical
value. Our theoretical picture is compared with experimental data
in Fig.~\ref{Fig:quantres}(a), where experimental and numerical
momentum distributions are shown at the quantum resonance
$\tau\simeq 2\pi$ ($T = 66.5\mu$s), after $N=30$ kicks. Note that
our numerical simulation exhibits the ballistic wings of the
distribution, which are swamped by the noise background in the
experimental data. When processing such data, only momenta in the
window $[-60,60]$ were taken into account. Furthermore, the
experimental distribution exhibits an asymmetry around $n=0$ which
is not present in the theory. This is due to two effects: the
first, and most important, is that of the lock-in amplifier and
its associated low pass filter, used in the TOF measurement, which
slightly distort the momentum distribution. The second is that the
removal of gravity's effect by the crystal phase modulator is
imperfect; gravity breaks the symmetry of the system's evolution
and hence of the momentum distribution. Nevertheless, the
experimental and theoretical results agree very well in the
central part of the distribution. Other deviations of the
experimental system from the ideal are: (i) pulses are not
$\delta$-like as they have a finite duration $t_{p}$, (ii) random
amplitude noise is introduced by laser power fluctuations ($\pm
5$\%), and (iii) different atoms are subject to somewhat different
values of $\phi_d$ \cite{darcy2001a}.

The addition of noise reshuffles the quasimomenta of the initial
distribution, at a rate proportional to ${\overline
n}_{\mbox{\tiny SE}}$, and thus destroys the conservation of
quasimomentum. This reshuffling prevents atoms from remaining in
the fast-travelling quasimomentum range for a long time; the
formation of ballistic wings is thus inhibited. On the other hand,
reshuffling gives {\em all}\/ atoms a chance of sojourning a while
in those quasimomentum ranges and hence experiencing a transient
ballistic momentum growth. As a result, the distribution at
moderate momenta broadens in time, at the expense of the ballistic
wings. This is seen in Fig.~\ref{Fig:quantres}(b): for ${\overline
n}_{\mbox{\tiny SE}}\simeq 0.14$, the momentum distribution is
strongly broadened as it exhibits enhanced population of moderate
momentum states.

The incoherent dynamics are amenable to analytical treatment
\cite{WGF2003}, which shows that the distribution no longer
approaches a stationary form. Instead, it evolves towards a
continually broadening, diffusion-like Gaussian. The
theoretically-obtained line shape compares very favorably to the
experimental one, as shown in Fig.~\ref{Fig:quantres}(b). Note
that we chose ${\overline n}_{\mbox{\tiny SE}}=0.1$ for the
analysis in order to achieve an optimal fit to the experimental
data. The uncertainty in the intensity of D2 light experienced by
the atoms, due to loss at the glass faces of the vacuum system and
the exact shape of the $2\mu$s pulse, means that the experimental
value of ${\overline n}_{\mbox{\tiny SE}}$ is $(0.14\pm 0.04)$,
consistent with this best theoretical fit.

Analytically, the mean energy grows according to $E \simeq
(D/2+\phi_{d}^{2}/4)N$, where $D\simeq {\overline n}_{\mbox{\tiny
SE}}/12$ is the diffusion coefficient associated with the momentum
transfer due to SE \cite{WGF2003}. Since $D$ is rather small
($\simeq 0.01$) for the cases considered here, the mean energy
growth is almost the same as in the resonant case {\em without}\/
decoherence, where the same expression for $E$ is obtained, except
$D=0$ (see above). Hence weak decoherence destroys the
conservation of quasimomentum, which lies at the very root of
resonances, yet in the ideal model it only affects the resonant
energy growth mildly. However, the theoretically almost identical
energy growth is produced by quite different physical mechanisms,
which react to the cutoffs inherent in experimental detection
schemes in dramatically different ways. In the coherently-evolved
case, the growth of the experimentally measured energy is strongly
depressed as soon as the ballistic wings escape the detection
windows. In the presence of noise, the energy growth is not due to
the {\em ballistic}\/ wings, but rather to {\em diffusive}\/
broadening of the whole distribution, and is therefore dominated
by the center. Hence the effect of finite experimental detection
windows is much less severe and the measured mean energy remains
closer to its ideal value up to a higher value of $N$. This leads
to an apparent enhancement of the resonance peaks compared to the
SE-free case. Such noise-induced signal enhancement is reminiscent
of `stochastic resonance', where the response of a system to some
input signal is enhanced by stochastic activation
\cite{bulsara96}. However, the hallmark of stochastic resonance is
a maximum signal enhancement at an optimal, nonvanishing noise
level. This has not so far been established in our present
scenario.

In summary, we have shown that the linear growth with time of the
mean energy at quantum resonance, inhibited in experiments by
finite detection windows on finite time scales, is restored there
by adding noise. This effect is ultimately rooted in the
difference between atoms and rotors and is a striking, albeit
indirect, demonstration of the peculiar nature of coherent
resonant transport for kicked atoms, and how it is modified by
photon recoil-induced decoherence.

We thank S.A. Gardiner and K. Burnett for stimulating and
enlightening discussions. We acknowledge support from the UK
EPSRC, The Royal Society, the EU TMR `Cold Quantum Gases' network
and QTRANS RTN1-1999-08400, the INFM-PA project {\it Weak Chaos:
theory and applications}, the US-Israel Binational Science
Foundation (BSF), and the Minerva Center of Nonlinear Physics of
Complex Systems.

\end{document}